# Rare coding SNP in DZIP1 gene associated with late-onset sporadic Parkinson's disease

André X. C. N. Valente[1,2,3], Joo H. Shin[4], Abhijit Sarkar[5] and Yuan Gao[4]

[1] Systems Biology Group, Biocant - Biotechnology Innovation Center, Cantanhede, Portugal
[2] CNC - Center for Neuroscience and Cell Biology, University of Coimbra, Coimbra, Portugal
[3] Center for the Study of Biological Complexity, Virginia Commonwealth University, Richmond, Virginia, USA
[4] Lieber Institute for Brain Development, Baltimore, Maryland, USA
[5] Physics Department, Catholic University of America, Wash8ington DC, USA

We present the first application of the hypothesis-rich mathematical theory (1) to genome-wide association data. The Hamza et al. (2) late-onset sporadic Parkinson's disease genome-wide association study dataset was analyzed. We found a rare, coding, non-synonymous SNP variant in the gene DZIP1 that confers increased susceptibility to Parkinson's disease. The association of DZIP1 with Parkinson's disease is consistent with a Parkinson's disease stem-cell ageing theory (1).

**Introduction**

Familial genetic linkage studies have associated six genes with Mendelian inheritable forms of Parkinson's disease (PD) (3; 4). However, these monogenic forms account for fewer than 10% of PD cases. Further, they lead mostly to juvenile or early onset forms of PD (before age 50). Given that no decisive environmental causative factors have been found either, the etiology of late-onset PD (comprising over 90% of all PD cases) remains essentially undetermined. A range of hypotheses are being explored (5; 6; 7; 8; 9; 10). We have proposed the theory that i) sporadic PD is best defined as a characteristic deviation from normality in the expression program of a cell (the PD-state) and ii) this PD-state can originate as a case of hematopoietic stem-cell program defect (1).

At present, considerable efforts are focused on finding differential genetic susceptibility to late-onset PD via the genome-wide association study (GWAS) (11). In a GWAS, a set of patients and controls is genotyped at known SNP sites in the human genome. Mathematically, this assigns individuals to locations within a high-dimensional SNP space (Figure 1). Genetic susceptibilities are inferred from statistically significant differences in the placement of patients and controls in this SNP space. Large enough differential disease risks constitute practical predictive genetic markers. So far though, susceptibilities found have been typically weak (some 85% of trait associated SNPs reported have an odds ratio in the 0.5-2 range) (12). Nonetheless, such findings can still be invaluable as indicators of the involvement of particular genes or biological processes in the disease mechanics. As of today, GWASs have reported about a dozen, modest effect (odds ratio in the 0.5-2 range), susceptibility loci for PD (2; 13; 14; 15; 16; 17; 18; 19; 20; 21).

The hypothesis-free paradigm currently dominates GWAS statistical data analysis (11). We have previously described why this is a poor choice (1). Biological knowledge and insightful hypotheses are as crucial in the analysis of a GWAS as they are in the analysis of any classical biological experiment. The alternative hypothesis-rich mathematical theory recognizes this fact and allows biological thought to maximize statistical power (1). Key in the approach is the concept of Rational Class (RC), a set of candidate laws (markers in the GWAS context) that share an underlying common rationale.

In this article, we analyze the late-onset sporadic PD GWAS dataset of Hamza et al. (2), under the hypothesis-rich framework (the late-onset, sporadic qualifier will be henceforth subsumed) (1; 22). In the *Methods* section, the focus is on describing the RCs constructed specifically for this PD GWAS analysis. Findings are summarized in the *Results* section. Finally, in the *Discussion* section, we review relevant biological information to contextualize our findings.

**Methods**

We analyzed the GWAS dataset of Hamza et al. (2), consisting of 1986 control subjects and 2000 sporadic late-onset PD patients. All individuals were Americans of self-reported European ancestry. As in any GWAS, a concern is the presence of population structure in the cohort data (23). Likely the European population, due to historical and geographical factors, does not constitute, mating-wise, a single uniformly mixed population. Now, suppose the existing subpopulations have distinct susceptibilities to PD. This could be due to differences in genetic background, culture (e.g., diet), or physical environment. Regardless, a genetic marker of a subpopulation (e.g., a SNP variant typical of a subpopulation) would then effectively mark a distinct susceptibility to PD. This poses a problem, in that we would like to interpret markers as having a causative effect on PD susceptibility, which clearly would not be the case here. The issue may also arise merely by study recruitment centers in areas with distinct subpopulations not enrolling identical ratios of PD to control subjects. Note that although for simplicity we allude above to discrete subpopulations, generically the mixing makeup will have a continuous character.

To analyze the dataset of Hamza et al., we used the SNP space coordinate system shown in Figure 1. The relative overall location of individuals in SNP space (Euclidean distance wise) reflects the cohort population structure. Namely, relative locations are consistent with parental country of origin for those subjects that reported such information and for whom both parents had a common origin. This is visually clear upon a change of coordinates to principal component coordinates (Figure 2).

A variety of methods exist for mitigating the population structure problem (23; 24; 25). We chose to homogenize the population regarding the PD to control subject ratio, via individual weight knock-down. Briefly, this involves reducing the statistical weight of selected individuals to locally level the ratio of PD to control subjects throughout SNP space. A separate article describes in detail both the method and its application to this particular dataset (26). The homogenization procedure reduced the dataset to a net weight of 1904 PD patients and 1802 controls (a 7% size reduction). This will be the reference dataset henceforth.

We utilized the hypothesis-rich framework to investigate the dataset (1; 22). The hypothesis-rich framework provides a *targeted search, unbiased assessment* approach to the analysis of GWAS data. The *targeted search* assertion follows from biological considerations guiding the statistical search for genetic susceptibility factors. Specifically, biological information enters the mathematical analysis via the concept of Rational Class (RC), a set of candidate genetic markers that share a common rationale. Yet, in spite of the biased search, an *unbiased assessment* is obtained from a proper mathematical treatment of multiple hypotheses testing (1; 22).

We now describe the RCs constructed for the PD GWAS problem. Throughout, recall that separating markers into distinct RCs can be statistically advantageous if the resulting RCs have different True Quality Distribution and Correlation Structures (shorthand, TQDs) (1). This can be the case whenever a biological rationale underlies the marker separation. On the other hand,

RCs must be rank ordered and statistical resolution decreases with increasing rank, thus overly liberal RC creation is pointless (1; 22).

A total of 23 RCs were constructed (Table 1). The first 15 RCs, containing individual SNPs, were based on the following factors:

Genomic region - we grouped SNPs by whether they fell in a coding region, in the UTR or in the remainder of the genome. Confirming the distinct biological roles of these regions, past GWASs show the incidence of trait associated SNPs in them is not uniform (11).

SNP allele frequencies - These frequencies are affected by the degree of selective pressure on the associated haplotypes. Thus, on average, the character of SNPs with different allele frequency ratios may be distinct. We divided SNPs into three broad groups, based on their minor variant frequency: <10%, 10-30% and >30%. Also, note that we are comparatively more interested in larger odds ratio markers. Given two SNP markers showing the same statistical significance (ordinarily, same p-value), the one with the lower minor variant frequency necessarily shows a larger differential trait susceptibility (larger odds ratio). Thus, as an additional benefit, the above frequency breakdown effectively protects the search for rare variant, high odds ratio markers.

Hematopoietic fingerprints - Given our PD working theory, the set of SNPs occurring in genes with a function in the hematopoietic system acquires particular relevance. We recorded 2253 SNPs spread across 662 so called hematopoietic fingerprint genes (27). The genes were identified by Chambers et al. via global gene expression profiling of murine hematopoietic stem cells and their major differentiated lineages (NK-cells, T-cells, B-cells, monocytes, neutrophils and nucleated erythrocytes) (27).

Combination of the above factors yielded RCs 1 thru 15 (Table 1). In these RCs, SNPs were tested for association with differential PD risk three separate times, each time based on a different mode of splitting the SNP space (Figure 3, 1-dimensional split modes). The dominant and recessive modes were motivated by their well known biological counterparts. However, a situation where phenotype is significantly more assured only under homogeneous alleles is also biologically plausible. The extreme mode accommodates these cases by excluding individuals with heterogeneous alleles from the statistical comparison. In every case, the null hypothesis was that the two regions defined by the split mode present no different susceptibility to PD. Statistical comparison between the two chosen regions was done via the Fisher exact test.

RCs 16 thru 19 were based on SNP pairs. Given there are on the order of $10^6$ SNPs, potential SNP pairs are on the order of $10^{12}$. A RC containing such a large number of entries is unlikely to have a favorable TQD (1). It is therefore fundamental to prioritize SNP pairs. We generated one list of SNP pairs based on protein-protein physical interactions (28; 29; 30). For every two interacting proteins on different chromosomes, all SNP pairs with one SNP in each of the interacting proteins respective coding gene region were added to the list. The exclusion of protein pairs on the same chromosome excludes pairs of SNPs potentially in linkage disequilibrium. Protein-protein interactions were obtained from HPRD (~39000 interactions) (31). The SNP pairs were tested for association with differential PD risk five times, each time based on a different mode of splitting the SNP space (Figure 3, 2-dimensional split modes). In every case, the null hypothesis was that the two regions defined by the split mode present no different susceptibility to PD. Statistical comparison between the two chosen regions was done via the Fisher exact test. The results of the tests were assigned to RC 16 or to RC 17 depending on whether the associated odds ratio was smaller or larger than 3. Once more, this has the benefit of safeguarding the search for high odds ratio markers.

A second list of SNP pairs was constructed based on the hematopoietic fingerprint genes. Based on the expression profiling, Chambers et al. had further divided the hematopoietic fingerprint genes into the following subclasses: hematopoietic stem cells, B-cells, naive T-cells, NK-cells, monocytes, granulocytes, nucleated erythrocytes, differentiated shared fingerprint, lymphoid shared fingerprint and myeloid shared fingerprint (27). We generated hematopoietic gene pairs by considering every possible pairing of genes within the same hematopoietic subclass, exclusive of gene pairs in the same chromosome. The procedure described above for protein pairs was then applied to the hematopoietic gene pairs, thus generating RCs 18 and 19.

RCs 20 thru 22 contained SNP tuplets generated from protein complexes (32; 33). Human protein complexes were obtained from the CORUM database (~1300 complexes) (34). Consider first RC 20, containing 2-tuplets generated from complexes of up to 4 proteins. The 2-tuplets for RC 20 were generated as follows:

   1. Given a complex, consider the SNPs that fall in the coding region of the protein members of the complex. Denote them as CSNPs. Add every possible (CSNP A, CSNP B) 2-tuplet to RC 20, provided CSNP A and CSNP B are located in different chromosomes.

   2. Repeat for every complex of up to 4 proteins.

Each 2-tuplet was tested for association with differential PD risk 3 separate times, as follows:

   1. Under the dominant 1-dimensional split mode, assign a Fisher exact test based p-value to each CSNP in the tuplet in the standard fashion (i.e., considering the CSNP as an individual SNP, as in the RCs 1 thru 15). We formalize it by writing p-value = p(CSNP; dominant mode).

   2. The p-value associated with the 2-tuplet is $(max(p(CSNP\ A;\ dominant\ mode), p(SNP\ B;\ dominant\ mode)))^2$ (i.e., squared).

   3. Assign two more p-values to the tuplet, as above, but now utilizing the recessive and extreme 1-dimensional split modes.

RC 21 was similar to RC 20, except that:

   1. It was based on complexes of sizes 3 thru 9.

   2. It contained 3-tuplets (CSNP A, CSNP B, CSNP C).

   3. The p-value associated with a 3-tuplet is $(max(p(CSNP\ A;\ split\ mode), p(SNP\ B;\ split\ mode), p(SNP\ C;\ split\ mode)))^3$ (i.e., cubed).

RC 22 was similar to RC 20, except that:

   1. It was based on complexes of sizes 4 thru 16.

   2. It contained 4-tuplets (CSNP A, CSNP B, CSNP C, CSNP D).

   3. The p-value associated with a 4-tuplet is $(max(p(CSNP\ A;\ split\ mode), p(SNP\ B;\ split\ mode), p(SNP\ C;\ split\ mode), p(SNP\ D;\ split\ mode)))^4$ (i.e., to the fourth power).

In these complex based RCs, in every case the null hypothesis is that *none* of the SNPs in the tuplet shows differential susceptibility to PD between the two regions defined by the split mode. The anticipation is that a complex mechanistically involved in PD may produce a tuplet (or tuplets) of particular low p-value under the above tuplet p-value definition. RCs 20, 21 and 22 are kept separate to preserve potentially distinct TQDs.

Finally, RC 23 was based on genes in the blood gene expression signature for PD (involving 18 genes) we developed in earlier work (1). RC 23 contained:

  1. All SNPs in the coding or UTR regions of the genes present in the expression signature.

  2. All SNP pairs, exclusive of pairs in the same chromosome, with one SNP in the coding or UTR regions of one expression signature gene and the other SNP in the coding or UTR region of a second expression signature gene.

The individual SNPs were tested for association with differential risk of PD under the three 1-dimensional split modes via the Fisher exact test, as in RCs 1 thru 15. The SNP pairs were tested for association with differential risk of PD under the five 2-dimensional split modes via the Fisher exact test, as in RCs 16 thru 19. All tests were placed in a single RC, given their low number.

**Results**

The significant findings from the hypothesis-rich analysis of the Hamza et al. dataset are presented in Table 2. For these SNPs, the null hypothesis was that the two regions defined by the *split mode* (see Figure 3: 1-dimensional split modes) present no differential susceptibility to PD. Now, pure chance in the finite sampling of individuals from the population could create the false impression of differential susceptibility. The *reference probability* indicates the likelihood that such stochasticity lead to an unwarranted call (as per the hypothesis-rich framework) of differential susceptibility (22). For easiness of comparison, the arbitrariness in defining *odds ratio* (given the validity of the inverse of any choice) was settled by making every odds ratio larger than unity. The *minor allele effect* entry then indicates which region carries the greater risk of PD.

The findings of relevant SNPs in the SNCA region, in the HLA-DRA region and in the chromosome 17 q21.31 region (usually categorized as the MAPT region) confirm previous GWASs observations (albeit, the diversity of SNPs found in this latter region deserves special attention) (2; 13; 14; 15; 16; 17; 18; 19; 20; 21). The novel finding is the increased susceptibility to PD conferred by a rare, coding, non-synonymous SNP variant in the DZIP1 gene (Figure 4).

**Discussion**

The PD working theory we put forward in previous work (1) provides a possible context for the connection of DZIP1 with PD found in this analysis. Therefore, we start by reviewing it. Firstly, PD would be defined in terms of the PD-state, a characteristic deviation from the normal expression program of a healthy cell. Singular cellular manifestations of PD would therefore be de-emphasized in favor of this systems-level definition. Crucially, the PD-state would be a generic cell state, not restricted to neurons. Secondly, the PD-state would originate in a stem-cell program defect, associated with the ageing of stem-cells. We proposed the hematopoietic stem-cell niche as a plausible place of origin for the PD-state, although other stem cell niches should not be ruled out from playing a part. Thirdly, the subsequent PD-state propagation to other cells would not occur evenly. Propagation would be faster to cells more amenable to reprogramming (such as other stem cells or their not yet fully differentiated progeny). Thus tissues under active regeneration would be the first to be affected. Beyond PD biology, note the validity of this theory would signal an effective degree of communication between different stem-cell niches greater then what is currently accepted.

We now describe what is known at present about the biological role of DZIP1. The gene DZIP1 encodes a C2H2-type zinc finger protein (35). Its acronym stands for DAZ-interacting protein 1, as DZIP1 was originally identified in a screen for protein interaction partners of the DAZ (deleted in azoospermia) protein (35). Its expression in human embryonic, stem, fetal and adult germ cells was thus well noted (35). Zebrafish mutants in *iguana* (the DZIP1 ortholog in Zebrafish) have been invaluable in characterizing the gene. A *iguana* mutant (*fo10a*) displayed ultrastructural defects in perivascular mural cell recruitment and subsequent hemorrage, thus linking vascular stability and DZIP1 (36). Work with Zebrafish *iguana* mutants also revealed DZIP1 to be a component of the Hedgehog (Hh) signaling pathway (37; 38). Within the Hh pathway, DZIP1 acts downstream of Smoothened, modulating the activity of the Gli family of transcription factors (37; 38). DZIP1 has further been implicated in primary ciliogenesis and its role in Hh signaling may occur in this context (39; 40; 41). Hh plays a vital part in directing embryonic pattern formation (42). However, it continues regulating adult stem cells beyond embryogenesis (43; 44). Studies have specifically implicated Hh in the adult maintenance of hematopoietic stem cells (45), epithelial stem cells in the gastrointestinal tract (46), neuronal stem cells in the subventricular zone (SVZ) and the hippocampal dentate gyrus (47; 48), hair follicle stem cells (49), mammary stem cells (50) and mesenchymal stem cells (51). Besides its role in neurogenesis, Hh has also shown neurotrophic properties, in particular regarding dopaminergic neuron survival (52; 53; 54). Administration of Sonic Hedgehog reduced behavioral deficits in animal models of PD (55; 56). Nonetheless, an earlier targeted genetic analysis of Sonic Hedgehog in Parkinson's patients, did not find any significant mutations in this gene (57).

Genetic mutations affecting the Hh pathway have been associated with an increased incidence of a diversity of cancers (see Merchant et al. (58) or Beachy et al. (44) for comprehensive listings). Under a cancer stem-cell hypothesis (59; 60) interpretation, this is consistent with the role of Hh in adult stem cell homeostasis. The aberrant Hh signaling would contribute the conversion of adult stem cells (or perhaps their early progeny) into cancer stem cells, cells endowed with stem-cell properties and trapped in a pathological state of constant renewal (44; 59). Now, under our PD hypothesis, PD also originates in a stem-cell program defect. However, while in the cancer stem cell hypothesis the pathology progresses via physical replication of the cancer stem cells themselves, in PD we are proposing propagation solely of the PD characteristic expression state (the PD-state) (1). The PD-state of a cell could possibly be physically locked in by epigenetic DNA modifications (1).

**Notes**

Quality control

At the outset, a quality control procedure (61) was applied to the Hamza et al. dataset that excluded the following SNPs from the entire analysis:

- SNPs with a p-value less than $10^{-5}$ under the Hardy-Weinberg test.
- SNPs with less than a 99.9% call rate.

The quality control was implemented using the program Plink (62). A total of 748807 SNPs passed the quality control.

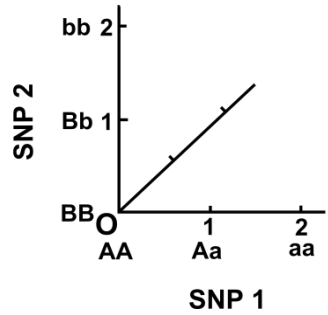

**Figure 1.** In a genome-wide association study (GWAS), subjects are vectors in SNP space. Depicted is one sensible coordinate system for SNP space. Capital letters represent the major allele, lower case letters the minor allele. To each SNP therefore corresponds an axis with 3 admissible values (0, 1 and 2). At present, typical cohort sizes are in the range of $10^3$ to $10^4$ subjects, while the number of SNPs genotyped is on the order of $10^6$.

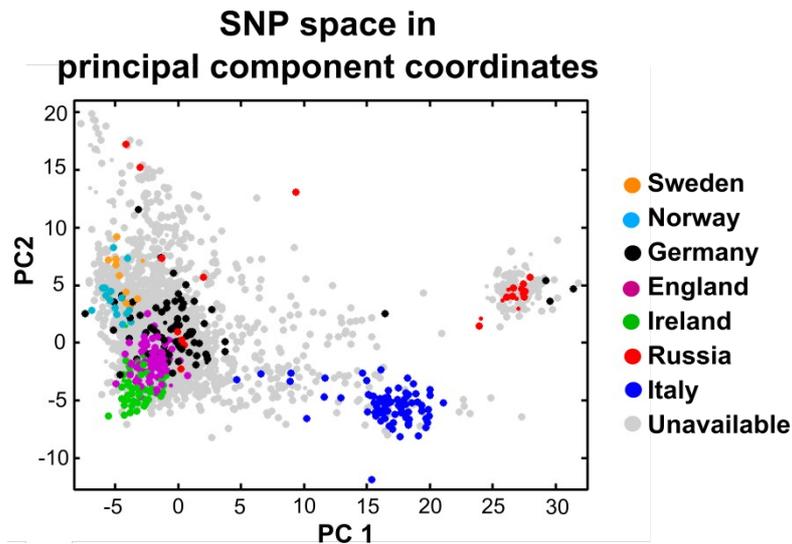

**Figure 2.** The Hamza et al. (2) cohort data in SNP space, after a change from the Figure 1 coordinate system to principal component coordinates (first two principal components shown). Color indicates the country of parental origin for subjects that reported such information and for whom both parents had a common origin. The plot replicates a similar figure in Hamza et al.. Smaller circles denote individuals with a lower statistical weight, due to the process of population homogenization across SNP space regarding the PD to control subject ratio (26).

| Rational Class Rank | Rational Class Description |
|---|---|
| 1 | **coding** \| allele freq.: < 10% \| dominant, recessive and extreme |
| 2 | **coding** \| allele freq.: 10% to 20% \| dom., rec. extr. |
| 3 | **coding** \| allele freq.: > 30% \| dom., rec. extr. |
| 4 | **UTR** \| allele freq. < 10% \| dom., rec. extr. |
| 5 | **UTR** \| allele freq.: 10% to 20% \| dom., rec. extr. |
| 6 | **UTR** \| allele freq.: > 30% \| dom., rec. extr. |
| 7 | **hematopoietic** \| coding \| allele freq. < 10% \| dom., rec. extr. |
| 8 | **hematopoietic** \| coding \| allele freq.: 10% to 20% \| dom., rec. extr. |
| 9 | **hematopoietic** \| coding \| allele freq.: > 30% \| dom., rec. extr. |
| 10 | **hematopoietic** \| UTR \| allele freq. < 10% \| dom., rec. extr. |
| 11 | **hematopoietic** \| UTR \| allele freq.: 10% to 20% \| dom., rec. extr. |
| 12 | **hematopoietic** \| UTR \| allele freq.: > 30% \| dom., rec. extr. |
| 13 | **non-coding/non-UTR** \| allele freq.: > 30% \| dom., rec. extr. |
| 14 | **non-coding/non-UTR** \| allele freq. < 10% \| dom., rec. extr. |
| 15 | **non-coding/non-UTR** \| allele freq.: 10% to 20% \| dom., rec. extr. |
| 16 | **protein pairwise interactions** \| coding \| OR > 3 \| 5 split modes |
| 17 | **protein pairwise interactions** \| coding \| OR < 3 \| 5 split modes |
| 18 | **hematopoietic pairwise interactions** \| coding \| OR > 3 \| 5 split modes |
| 19 | **hematopoietic pairwise interactions** \| coding \| OR < 3 \| 5 split modes |
| 20 | **protein complexes** \| sizes 2 thru 4 \| top 2 proteins |
| 21 | **protein complexes** \| sizes 3 thru 9 \| top 3 proteins |
| 22 | **protein complexes** \| sizes 4 thru 16 \| top 4 proteins |
| 23 | **gene expression sig.** \| coding and UTR \| all freqs. \| dom., rec., extr. |

**Table 1.** The Rational Classes (RCs) constructed to analyze the PD GWAS data.

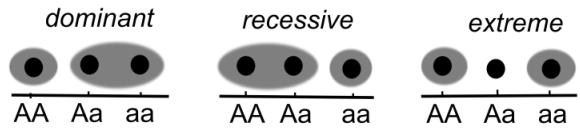

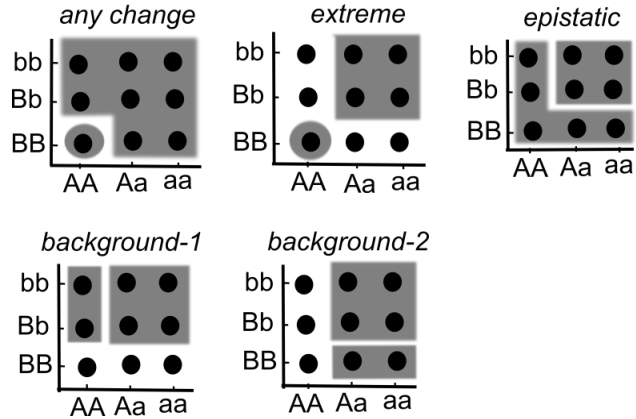

**Figure 3.** Each graph shows a manner of splitting SNP space into two shaded regions. Differential risk of PD between the shaded regions is then ascertained (non-shaded regions are ignored). **1-dimensional split modes:** Utilized in RCs containing single SNPs (RCs 1 thru 15 and RC 23). **2-dimensional split modes:** Utilized in RCs containing pairs of SNPs (RCs 16 thru 19).

| Rational Class | RC #2 coding region minor allele freq: 10%-30% | RC #7 hematopoietic coding region minor allele freq.< 10% | | RC #15 generic (non-coding/non-UTR) 30% < minor allele freq. |
|---|---|---|---|---|
| Gene | **IMP5, MAPT, CRHR1, KIAA1267, C17orf69, NSF** | DZIP1 | HLA-DRA | SNCA-GPRIN3, SNCA |
| SNP | rs12373123, rs12185235, 17651549, rs16940665, 12185268, kgp6408681, rs1052551, rs16940674, rs36076725, rs17652121, kgp3974170, rs17574604, kgp3365508, rs10445337, rs1881193, rs3583914, rs1052553, kgp4886152, rs199533 | kgp1112497 | rs3129822 | rs356220, rs2736990, rs356168 |
| Location | chr. 17, q21.31 in coding regions synonymous & non-synonymous substitutions | chr. 13, q32.1 coding region non-synonymous substitution | chr. 6, p21.3 intronic | chr. 4, q22.1 intergenic (SNCA-GPRIN3) intronic (SNCA) |
| Minor allele frequency | 19% thru 21% | 0.7% | 44% | 40% thru 49% |
| Minor allele effect | protective | harmful | harmful | harmful |
| Split mode | minor allele dominant | minor allele dominant | extreme | minor allele dominant |
| Odds ratio | 1.2 thru 1.3 | 4.4 | 1.5 | 1.3 thru 1.4 |
| Reference probability | 0.06 thru 0.09 | 0.03 | 0.04 | 0.04 thru 0.07 |

**Table 2.** Summary of the findings from hypothesis-rich analysis of the Hamza et al. GWAS PD dataset. See the *Results* main text section for meaning of the entries.

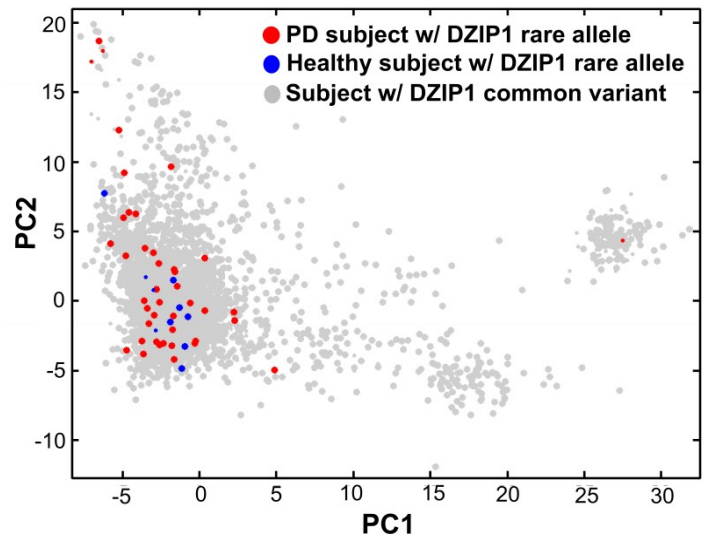

**Figure. 4.** Individuals in the Hamza et al. cohort carrying a copy of the rare DZIP1 allele are highlighted in SNP space, under principal component coordinates (first two principal components shown). No homozygous rare allele individuals were present in the dataset.